\documentclass[journal]{IEEEtran}
\usepackage[pdftex]{graphicx}
\graphicspath{{../pdf/}{../jpeg/}}
\DeclareGraphicsExtensions{.pdf,.jpeg,.png}
\usepackage[cmex10]{amsmath}
\usepackage{amsfonts}
\usepackage{amsthm}
\usepackage[english]{babel}
\usepackage{graphicx}
\usepackage{cite}
\usepackage{blindtext}
\usepackage{tcolorbox}
\usepackage{xcolor}
\usepackage{lipsum}
\usepackage{algorithm}
\usepackage{algorithmic}
\usepackage{array} 
\usepackage{tabularx}
\usepackage{balance}
\usepackage{makecell}
\usepackage{subcaption}

\usepackage[T1]{fontenc} 

\newtheorem{proposition}{Proposition}
\newtheorem{lemma}{Lemma}
\usepackage{amssymb}
\begin{document}
	
	\title{\huge Multi-Objective Resource Allocation for IRS-Aided SWIPT}
	
	\author{Ata Khalili,~\textit{Member,~IEEE},~Shayan Zargari,~Qingqing Wu,~\textit{Member,~IEEE},\\~Derrick Wing Kwan Ng,~\textit{Fellow,~IEEE},~and Rui Zhang,~\textit{Fellow,~IEEE}
		\vspace{-10mm}
		\thanks{ Ata Khalili is with Electronics Research Institute,~Sharif University of Technology, Tehran,~Iran and also with the Department of Electrical and Computer Engineering Tarbiat Modares University,~Tehran, Iran (e-mail: ata.khalili@ieee.org).~Q. Wu is with the State Key Laboratory of Internet of Things for Smart City, University of Macau,~999078, and also with the National Mobile Communications Research Laboratory, Southeast University, Nanjing 210096, China.~The work of Q. Wu is supported in part by FDCT 0108/2020/A, and the Open Research Fund of National Mobile Communications Research Laboratory, Southeast University (No. 2021D15).~D.~W.~K.~Ng is with the School of
			Electrical Engineering and Telecommunications,~University of New South
			Wales,~Sydney,~NSW 2052,~Australia.~D. W. K. Ng is supported by funding from the UNSW Digital Grid Futures Institute, UNSW, Sydney, under a cross-disciplinary fund scheme and by the Australian Research Council's Discovery Project (DP210102169).~Rui Zhang is with the Department of Electrical and
			Computer Engineering,~National University of Singapore,~Singapore 117583.}}   
	\maketitle
	
	\begin{abstract}
		In this letter,~we study the resource allocation for a multiuser intelligent reflecting surface (IRS)-aided simultaneous wireless information and power transfer (SWIPT) system.~Specifically,~a multi-antenna base station (BS) transmits energy and information signals simultaneously to multiple energy harvesting receivers (EHRs) and information decoding receivers (IDRs) assisted by an IRS.~Under this setup,~we introduce a multi-objective optimization (MOOP) framework to investigate the fundamental trade-off between the data sum-rate maximization and the total harvested energy maximization,~by jointly optimizing the energy/information beamforming vectors at the BS and the phase shifts at the IRS.~This MOOP problem is first converted to a single-objective optimization problem (SOOP) via the $\epsilon$-constraint method and then solved by majorization minimization (MM) and inner approximation (IA) techniques.~Simulation results unveil a non-trivial trade-off between the considered competing objectives,~as well as the superior performance of the proposed scheme as compared to various baseline schemes.
	\end{abstract}
	
	\vspace{-6mm}
	\section{Introduction}
	\vspace{-2mm}
	Simultaneous wireless information and power transfer (SWIPT) has been introduced as a promising solution for addressing the energy limitation of battery-powered devices,~especially in low-power Internet-of-Things (IoT) scenarios.~In such scenarios,~the performances of wireless communication and energy transfer are both important,~which motivates the investigation on their fundamental trade-off \cite{Zhou}.~As a result,~multi-objective optimization problem (MOOP) has been proposed to address conflicting objectives in SWIPT systems.~For instance,~\cite{Zlatanov} studied the trade-off between the transmit power and the total harvested power in a full-duplex SWIPT system.
	
Recently,~intelligent reflecting surface (IRS) been proposed as a promising solution to improve the wireless communication spectral and energy efficiency,~which has been considered for the future sixth-generation (6G) wireless network \cite{Zhang}.~In general,~an IRS is composed of a large number of reconfigurable passive reflecting elements installed on a planar surface and each reflecting element can introduce a phase shift when reflecting the incident signal.~By properly designing the phase shifts of these elements,~the reflected signals from the IRS can be combined at the receivers either constructively or destructively to improve the desired signal power or to eliminate the undesired interference \cite{Wu2}.~In \cite{Schoberr}, an IRS-aided green system was considered where the total transmit power was minimized by jointly optimizing the passive and active beamformers based on the alternative optimization (AO) and inner approximation (IA) methods.~Existing works have provided researchers with clear evidence that the deployment of IRSs can significantly improve the performance of wireless communication systems.~However,~they mostly focus on pure wireless information transmission while its potential performance gain to wireless power transfer remains unclear.~In this regard,~some recent works have exploited the IRS in SWIPT systems, e.g., \cite{Wu6,Wu5,Tang,Ata_WCL2}.~In \cite{Wu6}, the total transmission power was considered in a MISO IRS-aided SWIPT network where an AO algorithm via applying the penalty-based method was adopted. The authors in \cite{Wu5} aimed to maximize the weighted sum-power in the formulation of single-objective optimization problem (SOOP) subject to the individual SINR constraints at the IDRs.~In particular, they proposed an AO method to obtain a suboptimal solution of the main problem by optimizing the IRS phase shifts and transmission precoders at the base station (BS) iteratively.~Compared with existing works, the main contributions of this letter are summarized as follows:

	$\bullet$ We propose an IRS-SWIPT system in which both information decoding receivers (IDRs) and energy harvesting receivers (EHRs) receive signals reflected by the IRS such that the overall performance in terms of data sum-rate and total harvested power can be improved.~To balance between the information transmission and energy harvesting (EH) in the IRS-SWIPT system,~we formulate an MOOP by optimizing joint active and passive beamforming at the BS and IRS,~respectively.~In contrast to \cite{Yu,Wu2,Wu6,Wu5,Tang,Ata_WCL2},~where only one of the two objectives was investigated with the other being fixed,~we propose a new MOOP framework to fundamentally characterize their trade-off.
	
	$\bullet$ The resulting non-convex MOOP is transformed into a single-objective optimization problem (SOOP) via the $\epsilon$-constraint method,~which is more efficient than the conventional weighted sum maximization method as the former can characterize the entire Pareto boundary of the trade-off region \cite{MOOP}.
	
	$\bullet$ Note that for existing methods in the literature, e.g., semi-definite relaxation (SDR) and AO based method \cite{Yu,Wu2}, \cite{Wu6,Wu5,Tang,Ata_WCL2}, the convergence is not always guaranteed due to the application of Gaussian randomization in solving the feasibility check problem. In contrast, via applying the IA method as in our work, obtaining a locally optimal solution is guaranteed \cite{Schoberr}.

	\textit{Notation:} Vectors and matrices are indicated by
	boldface lower-case letters and capital letters,~respectively.~For a square matrix $\mathbf{A}$,~$\mathbf{A}^H$,~$\mathbf{A}^T$,
	$\text{Tr}(\mathbf{A})$,~$||\mathbf{A}||_{*}$,~and $\text{Rank}(\mathbf{A})$ denote its Hermitian conjugate transpose,~transpose,~trace,~trace norm,~and rank,~respectively.~$\mathbf{A}\succeq\mathbf{0}$ means that $\mathbf{A}$ is a positive semidefinite matrix.~$\text{diag}(\cdot)$ is the diagonalization operation.~The Euclidean norm of a complex vector and the absolute value of a complex scalar are denoted by $\|\cdot\|$ and $|\cdot|$,~respectively.~$\nabla_{\mathbf{x}}f(\mathbf{x})$ denotes the gradient vector  with respect to $\mathbf{x}$.~The expectation operator is denoted by $\mathbb{E}[\cdot]$,~and $\mathbb{C}^{M\times N}$ represents $M\times N$ dimensional complex matrices.~The distribution of a circularly symmetric complex Gaussian (CSCG) random vector with mean $\boldsymbol{\mu}$ and covariance matrix $\mathbf{C}$ is denoted by $\sim \mathcal{C}\mathcal{N}(\boldsymbol{\mu},\,\mathbf{C})$.
	\begin{figure}[t]
		\centering
		\includegraphics[width=2.50in]{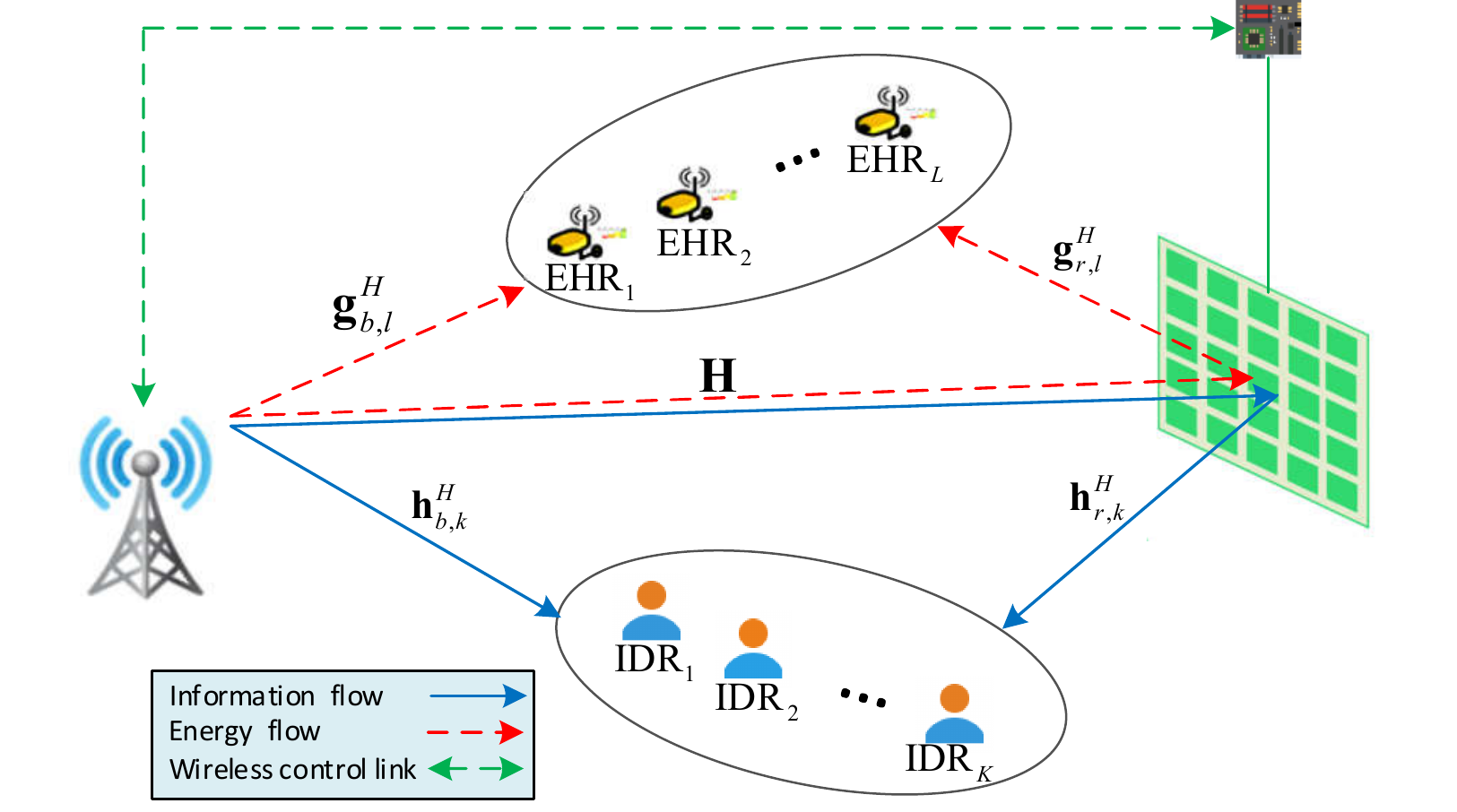}
		\caption{An IRS-aided multiuser MISO SWIPT system.}
		\vspace{-5mm}
	\end{figure}
	\vspace{-5mm}
	\section{System Model}
	\subsection{Signal Model}
	In this letter,~we consider an MISO downlink (DL) IRS-SWIPT system consisting of a BS,~an IRS,~$K$ IDRs,~and $L$ EHRs as shown in Fig.~1.~The BS and IRS are equipped with $M$ antennas and $N$ reflecting elements,~respectively.~It is assumed that all the receivers are single-antenna devices to reduce the hardware cost and complexity.~The transmit signal at the BS can be written as $\mathbf{s}=\sum_{k \in \mathcal{K}} {{\mathbf{w}_k}{x^{\text{ID}}_k}}+\sum_{l \in \mathcal{L}}\mathbf{v}_{l}x^{\text{EH}}_{l}$,~where $x^{\text{ID}}_{k}$ and $x^{\text{EH}}_{l} \in \mathbb{C}$ are the information signal for IDR $k\in\mathcal{K}=\{1,...,K\}$ and energy signal for EHR $l\in\mathcal{L}=\{1,...,L\}$, respectively.~Without loss of generality, ${x^{\text{ID}}_k}$ is assumed to be independent and identically distributed (i.i.d) while satisfying $\mathbb{E}\{|x^{\text{ID}}_k|^2\}=1$ and we further assume that $x^{\text{EH}}_{l}$ are independently generated from an arbitrary distribution with $\mathbb{E}\{|x^{\text{EH}}_l|^2\}=1$.~Besides,~${\mathbf{w}}_{k}\in \mathbb{C}^{M \times 1}$ denotes the transmit information beam for IDR $k$ and ${\mathbf{v}}_{l}\in \mathbb{C}^{M \times 1}$ is the transmit energy beam for EHR $l$.
	\vspace{-5mm}
	\subsection{Channel Model}
	Assume that all the channel links experience a quasi-static flat fading and accurate channel state information (CSI) can be obtained through an IRS controller\footnote{The results in this letter serve as theoretical performance upper bounds for IRS-aided SWIPT systems with imperfect CSI in practice.} (see \cite{Zhang}).~The received signal at the $k$-th IDR is given by ${y^\text{ID}_k} = {\mathbf{h}}_k^H{\mathbf{s}} + {n^\text{ID}_k},$
	$k\in\mathcal{K}$, where $n^\text{ID}_{k}\sim\mathcal{C}\mathcal{N}(0,\,\sigma_{k}^2)$ is the received noise at the $k$-th IDR with variance $\sigma_{k}^2$.~In particular,~${\mathbf{h}}_k^H \buildrel \Delta \over=\mathbf{h}^H_{b,k}+{\mathbf{h}}_{r,k}^H{\mathbf{\Theta}}{\mathbf{{\mathbf{H}}}}$ represents the equivalent channel from the BS-to-IRS and IRS-to-IDR $k$ along with the direct path from the BS-to-IDR $k$.~Especially,~${\mathbf{h}}_{r,k}\in \mathbb{C}^{N\times 1}$,~$\mathbf{h}_{b,k}\in \mathbb{C}^{M\times 1}$,~and ${\mathbf{H}}\in \mathbb{C}^{N \times M}$ are the channel vectors between the $k$-th IDR and IRS,~BS and IDR $k$,~and the equivalent channel matrix between the BS and IRS,~respectively.
	Besides,~${\mathbf{\Theta}}=\text{diag} (\alpha_{1} e^{j\beta_{1}},...,~\alpha_{N} e^{j\beta_{N}})$ denotes the reflection coefficient matrix at the IRS where $\beta_{n} \in (0,2\pi]$ and $\alpha_{n} \in [0,1]$,~$\forall n \in \{1,...,N\}$,~represent the phase shift and reflection amplitude of the $n$-th IRS element,~respectively.~We assume that all passive elements have an amplitude equal to one i.e.,~$|\alpha_{n}|^{2}=1$ as commonly adopted in literature,~e.g.,~[7]--[11].~Accordingly,~the received signal at the $l$-th EHR can be written as $
	{y^\text{EH}_l} = {\mathbf{g}}_l^H{\mathbf{s}} + {{z}^\text{EH}_l},$
	where ${z}^\text{EH}_l\sim \mathcal{C}\mathcal{N}(0,\,\delta^2_l)$ is the received noise with variance $\delta^2_l$.~Similarly,~we have ${\mathbf{g}}_l^H \buildrel \Delta \over =\mathbf{g}^H_{b,l}+ {\mathbf{g}}_{r,l}^H{\mathbf{\Theta}} {\mathbf{{\mathbf{H}}}}$,~where ${\mathbf{g}}_{r,l}\in \mathbb{C}^{N\times 1}$ denotes the channel vector between the $l$-th EHR and IRS,~and $\mathbf{g}_{b,l}\in \mathbb{C}^{M\times 1}$ represents the channel vector between the $l$-th EHR and BS.
	\vspace{-5mm}
	\subsection{Performance Metrics}	\vspace{-1mm}Without loss of generality,~we assume that a pseudo-random sequence generator provides $\mathbf{v}_{l}$ at the BS with a given seed,~which is known to all users.~Therefore,~all IDRs can eliminate the possible interference caused by the energy signal \cite{Xu}.~The received signal-to-interference-plus-noise ratio (SINR) at the $k$-th IDR can be written as
	\begin{equation}\label{sinr}
	\text{SINR}_k({\mathbf{w}}_k,{\mathbf{\Theta}})= \frac{{{{\left| {\mathbf{{\rm{ {\mathbf{h}}}}}}_{k}^H {\mathbf{{{w}}}}_k \right|}^2}}}{{\sum_{i \in \mathcal{K},i \ne k} {{{\left|  {\mathbf{{\rm{ {\mathbf{h}}}}}}_{k}^H {\mathbf{{{w}}}}_i \right|}^2} + \sigma _k^2} }}.
	\end{equation}
	On the other hand,~the total amount of harvested energy at EHR $l$ is considered to be linearly proportional to the received power \cite{Wu5},~which is given by 
	\begin{equation}\label{harvested power}
	{P_l}({\mathbf{w}}_k,{\mathbf{v}}_l,{\mathbf{\Theta}})=\eta_l~    \mathbb{E}\big\{\sum_{k\in\mathcal{K}} {{\left|  {\mathbf{{\rm{ {\mathbf{g}}}}}}_{l}^H {\mathbf{{{w}}}}_k \right|}^2}+\sum_{l \in \mathcal{L}}{{\left|  {\mathbf{{\rm{ {\mathbf{g}}}}}}_{l}^H {\mathbf{{{v}}}}_l \right|}^2}\big\},
	\end{equation}
	where $0 \leq \eta_l \leq 1$ is the energy conversion efficiency of the $l$-th EHR.~Note that the power of received noise is neglected in (\ref{harvested power}) as it is negligible for EH.
	\vspace{-4mm}
	\section{Problem Formulation and Solution}
	In this section,~we first formulate two SOOPs for the purposes of maximizing the data sum-rate and the total harvested energy,~respectively.~Then,~we introduce an MOOP framework to investigate the trade-off between these two competing objectives.
	\vspace{-4mm}
	\subsection{Optimization Problem Formulation}
	First,~the sum-rate maximization problem is formulated as follows,~which jointly optimizes the covariance matrix of the energy signal,~active beamforming vectors at the BS,~and phase shifts at the IRS.\\
	
	\textit{Problem 1: Sum-rate Maximization}
	\begin{subequations}
		\begin{align}
		\text{P1:} \:\: &\underset{{\mathbf{v}_{l}},{\mathbf{w}}_k,{\mathbf{\Theta}}} {\text{maximize}} \quad \sum_{k\in\mathcal{K}}\log_2\big(1+\text{SINR}_k({\mathbf{w}}_k,{\mathbf{\Theta}})\big) \\
		&\text{s.t.} \quad\sum_{k\in\mathcal{K}}{\left\| {\mathbf{w}}_k\right\|^2}+\sum_{l\in\mathcal{L}}{\left\| {\mathbf{v}}_l\right\|^2}\leq P_{\text{max}},\label{P1-0}\\
		&\quad \quad \text{SINR}_k({\mathbf{w}}_k,{\mathbf{\Theta}})\geq \Gamma_{\text{req},k},\: \forall k,\label{p1-1}\\
		&\quad\quad P_l({\mathbf{w}}_k,{\mathbf{v}_{l}},{\mathbf{\Theta}})\geq E_{\text{min},l},~ \forall l,\label{p1-2}\\
		&\quad\quad  {|\mathbf{\Theta}_{{nn}}|}=1,~ \forall n,\label{p1-5}
		\end{align}
	\end{subequations}
	where $P_\text{max}$ indicates the maximum transmit power at the BS.~Constants $\Gamma_{\text{req},k}\geq 0$ and $ E_{\text{min},l}\geq 0$ denote the minimum required SINRs for the IDRs and minimum harvested energy requirement for the EHRs,~respectively.~Constraint (3e) is imposed to guarantee that the IRS only introduces phase shifts to the impinging signals.~Similarly,~for the total harvested energy maximization problem,~we impose the same constraint set as for (P1)\footnote{It should be noted that although solving (P1) with different values of $E_{\min,l}$ and (P2) with different values of $\Gamma_{\text{req},k}$ can also help investigate the trade-off between ID and EH, the set of all Pareto optimal resource allocation policies cannot be obtained in general \cite{Zlatanov}.}.~Then,~the problem is formulated as follows.\\
	
	\textit{Problem 2: Total Harvested Energy Maximization}
	\begin{subequations}
		\begin{align}
		\text{P2:} \:\: &\underset{{\mathbf{v}_{l}},{\mathbf{w}}_k,{\mathbf{\Theta}}} {\text{maximize}} \quad \sum_{l\in \mathcal{L}} P_l({\mathbf{w}}_k,{\mathbf{v}_{l}},{\mathbf{\Theta}})\\
		&\text{s.t.} \quad \text{ (\ref{P1-0})--(\ref{p1-5})}.
		\end{align}
	\end{subequations}
	Subsequently,~the MOOP based on (P1) and (P2) is formulated in the next.
	
	\textit{Problem 3: Multi-objective Optimization}
	\begin{subequations}
		\begin{align}
		\text{P3:} \:\:
		&\text{Q1:}\:\: \underset{{\mathbf{v}_{l}},{\mathbf{w}}_k,{\mathbf{\Theta}}} {\text{maximize}} \quad  \sum_{k\in\mathcal{K}}\log_2\big(1+\text{SINR}_k({\mathbf{w}}_k,{\mathbf{\Theta}})\big)  \\ &\text{Q2:}\:\:\underset{{\mathbf{v}_{l}},{\mathbf{w}}_k,{\mathbf{\Theta}}} {\text{maximize}} \quad  \sum_{l\in \mathcal{L}} P_l({\mathbf{w}}_k,{\mathbf{v}_{l}},{\mathbf{\Theta}})\\
		&\text{s.t.} \quad \text{ (\ref{P1-0})--(\ref{p1-5})}.
		\end{align}
	\end{subequations}
	\vspace{-12mm}
	\subsection{Proposed Solution}
	To address the conflicting objectives in (P3),~we adopt the $\epsilon$-constraint method \cite{Khalili} by transferring (Q2) to the constraint set and setting (Q1) as the main objective function.~Bear in mind that the $\epsilon$-constraint method can generate the whole Pareto frontier \cite{MOOP} of the two optimal objective values by varying the value of $\epsilon$ and solving the corresponding optimization problem.~Thus,~for a given $\epsilon$,~the new optimization problem for obtaining a Pareto optimal solution can be written as
	\vspace{-2mm}
	\begin{subequations}
		\begin{align}
		\text{P4:} \:\:
		&\underset{{\mathbf{v}_{l}},{\mathbf{w}}_k,{\mathbf{\Theta}}} {\text{minimize}} \quad  -\sum_{k\in\mathcal{K}}\log_2\big(1+\text{SINR}_k({\mathbf{w}}_k,{\mathbf{\Theta}})\big)  \\
		&\text{s.t.} \quad  \sum_{l\in \mathcal{L}} P_l({\mathbf{w}}_k,{\mathbf{v}_{l}},{\mathbf{\Theta}})\geq\epsilon,~\label{P4}\\
		&\quad\quad \text{ (\ref{P1-0})--(\ref{p1-5})}.
		\end{align}
	\end{subequations}
	
	Constraint (\ref{P4}) states that the total EH amount is required to be greater than $\epsilon$.~It is evident that the value of $\epsilon$ determines the relationship of the solution of (P4) compared with that of  (P3).~Besides,~due to the existence of coupling between optimization variables,~i.e.~${\mathbf{w}}_k$ and $\mathbf{\Theta}$,~as well as the unit-modulus constraints in (3e),~it is generally difficult to obtain a globally optimal solution for (P4).~As an alternative,~we aim to design a computationally efficient algorithm to obtain a suboptimal solution to (P4).~To start with,~we first define $\mathbf{W}_{k}=\mathbf{w}_{k}\mathbf{w}_{k}^{H}$,~$\mathbf{V}_{l}=\mathbf{v}_{l}\mathbf{v}_{l}^{H}$,~${\boldsymbol{\theta}}=( e^{j\alpha_{1}},...,\:e^{j\alpha_{N}})^H\in\mathbb{C}^{N\times1}$, and $\mathbf{u}=[{\boldsymbol{\theta}}^T \: t]^T\in\mathbb{C}^{(N+1)\times1}$,~respectively.~Besides,~$t\in\mathbb{C}$ is a dummy variable with $|t|=1$.~By applying the transformation of semidefinite programming (SDP),~${\mathbf{U}}=\mathbf{u}\mathbf{u}^H\in\mathbb{C}^{(N+1)\times(N+1)}$ is introduced such that the matrix ${\mathbf{U}}$ is semi-definite and satisfies $\text{Rank}({\mathbf{{U}}})\leq1$.~Thus,~$|(\mathbf{h}^H_{b,k}+{\mathbf{h}}_{r,k}^H{\mathbf{\Theta}}{\mathbf{{\mathbf{H}}}})\mathbf{w}_i|^2$,~$|(\mathbf{g}^H_{b,l}+ {\mathbf{g}}_{r,l}^H{\mathbf{\Theta}} {\mathbf{{\mathbf{H}}}})\mathbf{w}_k|^2$,~and $|(\mathbf{g}^H_{b,l}+ {\mathbf{g}}_{r,l}^H{\mathbf{\Theta}} {\mathbf{{\mathbf{H}}}})\mathbf{v}_{l}|^2$ can be equivalently written as  $\text{Tr}(\mathbf{U}\mathbf{L}_k\mathbf{W}_i\mathbf{L}_k^H)=\text{Tr}(\mathbf{W}_i\mathbf{Z}_k)$,~$\text{Tr}(\mathbf{U}\tilde{\mathbf{L}}_l\mathbf{W}_k\tilde{\mathbf{L}}_l^H)=\text{Tr}(\mathbf{W}_k{\mathbf{X}}_l)$,~and $\text{Tr}(\mathbf{U}\tilde{\mathbf{L}}_l\mathbf{V}_{l}\tilde{\mathbf{L}}_l^H)=\text{Tr}(\mathbf{V}_{l}{\mathbf{Y}}_l)$,~respectively,~where $\mathbf{L}_k=[(\text{diag}({\mathbf{h}}_{r,k}^H)\mathbf{H})^T\:  \mathbf{ h}_{b,k}^*]^T$,~$\tilde{\mathbf{L}}_l=[(\text{diag}({\mathbf{g}}_{r,l}^H)\mathbf{H})^T\: \mathbf{ g}_{b,l}^*]^T$,~$\mathbf{Z}_k=\mathbf{L}_k^H\mathbf{U}\mathbf{L}_k$,~and ${\mathbf{X}}_l=\tilde{\mathbf{L}}_l^H\mathbf{U}\tilde{\mathbf{L}}_l$.~Unlike most of the existing works adopting AO which optimizes $\mathbf{W}_k$,~$\mathbf{V}_l$,~and $\boldsymbol{\theta}$ separately in an iterative manner,~we aim to optimize all variables jointly.~However,~the multiplication of two matrices poses a challenge in solving our problem.~By following \cite{Schoberr,Ata_WCl},~we further rewrite the related terms as
	\vspace{-1mm}
	{\small \begin{align}
		&\text{Tr}(\mathbf{W}_i\mathbf{Z}_k)=\frac{1}{2}\left\|\mathbf{W}_i+\mathbf{Z}_k\right\|_F^2-\frac{1}{2}\left\|\mathbf{W}_i\right\|^2_F-\frac{1}{2}\left\|\mathbf{Z}_k\right\|^2_F\triangleq\mathbf{A}_{k,i},\label{8}\\
		&\text{Tr}(\mathbf{W}_k\mathbf{X}_l)=\frac{1}{2}\left\|\mathbf{W}_k+\mathbf{X}_l\right\|_F^2-\frac{1}{2}\left\|\mathbf{W}_k\right\|^2_F-\frac{1}{2}\left\|\mathbf{X}_l\right\|^2_F\triangleq\mathbf{B}_{k,l},\label{7}\\
		&\text{Tr}(\mathbf{V}_{l}\mathbf{Y}_l)=\frac{1}{2}\left\|\mathbf{V}_{l}+\mathbf{Y}_l\right\|_F^2-\frac{1}{2}\left\|\mathbf{V}_{l}\right\|^2_F-\frac{1}{2}\left\|\mathbf{Y}_l\right\|^2_F\triangleq\mathbf{C}_{l}.\label{9}
		\end{align}}Hence,~(P4) is now in a more tractable form,~which is given by  
	\begin{subequations}
		\begin{align}
		\text{P5:} \:\:
		&\underset{{\mathbf{V}}_l,{\mathbf{W}}_k,{\mathbf{U}}} {\text{minimize}} \quad  -\sum_{k\in\mathcal{K}}\log_2\big(1+\frac{\mathbf{A}_{k,k}}{{\sum\limits_{i \in \mathcal{K},i \ne k} {\mathbf{A}_{k,i} + \sigma _k^2} }}\big)  \\
		&\text{s.t.} \quad  \eta_l\big( \sum_{l\in \mathcal{L}} \sum_{k\in \mathcal{K}} \mathbf{B}_{k,l}+ \sum_{l\in \mathcal{L}}\mathbf{C}_l) \geq\epsilon,~\label{P6}\\
		&\quad\quad \sum_{k\in \mathcal{K}}\text{Tr}(\mathbf{W}_k)+\text{Tr}(\mathbf{V}_{l})\leq P_{\text{max}},\label{P6-0}\\
		&\quad\quad \frac{\mathbf{A}_{k,k}}{{{\Gamma_{\text{req},k}}}} - \sum_{ i \ne k} \mathbf{A}_{k,i} \ge \sigma _k^2,\: \forall k,\label{p6-22c}\\
		&\quad\quad \eta_l \sum_{k\in \mathcal{K}} \mathbf{B}_{k,l}+\mathbf{C}_l  \ge    E_{\text{min},l},~ \forall l,\label{p6-2}\\
		&\quad\quad \mathbf{V}_{l}\succeq\mathbf{0},~ \mathbf{W}_k\succeq\mathbf{0},~ \forall k,~l\label{p6-5}\\ 
		&\quad\quad \text{Rank}(\mathbf{W}_k)= 1,~\forall k,~\quad \text{Rank}(\mathbf{U})= 1.\label{p6-6}
		\end{align}
	\end{subequations}
	However,~(\ref{8}),~(\ref{7}),~and (\ref{9}) are not concave functions.~To handle them,~we adopt the iterative MM method \cite{Babu} via the first-order Taylor approximation to establish the corresponding convex lower bounds.~Taking (7) as an example,~the term $F_1(\mathbf{W}_{i},\mathbf{Z}_{k})\triangleq\frac{1}{2}\left\|\mathbf{W}_i+\mathbf{Z}_k\right\|_F^2$ can be bounded by an affine function which is given by  
	\begin{align}
	F_1(\mathbf{W}_{i},\mathbf{Z}_{k})&\geq F_1(\mathbf{W}^{(i)}_{i},\mathbf{Z}^{(i)}_{k})\nonumber\\&+\text{Tr}(\nabla_{\mathbf{W}_{i}}^HF_1(\mathbf{W}^{(i)}_{i},\mathbf{Z}^{(i)}_{k})(\mathbf{W}_{i}-\mathbf{W}_{i}^{(i)}))\nonumber\\&+\text{Tr}(\nabla_{\mathbf{Z}_{k}}^HF_1(\mathbf{W}^{(i)}_{i},\mathbf{Z}^{(i)}_{k})(\mathbf{Z}_k-\mathbf{Z}^{(i)}_{k}))\nonumber\\&\triangleq \tilde{F}_1(\mathbf{W}^{(i)}_{i},\mathbf{Z}^{(i)}_{k}).\label{16}
	\end{align}
	Similar to (\ref{16}),~we have $F_2(\mathbf{W}_{k},\mathbf{X}_{l})\triangleq\frac{1}{2}\left\|\mathbf{W}_{k}+\mathbf{X}_l\right\|_F^2\geq\tilde{F}_2(\mathbf{W}^{(i)}_k,\mathbf{X}^{(i)}_{l})$ and ${F_3}(\mathbf{V}_{l},\mathbf{Y}_l)\triangleq\frac{1}{2}\left\|\mathbf{V}_{l}+\mathbf{Y}_l\right\|_F^2\geq\tilde{F}_3(\mathbf{V}^{(i)}_{l},\mathbf{Y}^{(i)}_l)$,~where $\{\mathbf{W}_k^{(i)},\mathbf{W}^{(i)}_i,\mathbf{Z}^{(i)}_k,\mathbf{X}^{(i)}_{l},\mathbf{V}_{l}^{(i)},\mathbf{Y}^{(i)}_l\}$ is the set of solutions obtained at the $i$-th iteration of the MM method.~Therefore,~lower bounds of (\ref{8}),~(\ref{7}),~and~(\ref{9}) are given by
	\begin{align}
	&\tilde{\mathbf{A}}^{(i)}_{k,i}\triangleq\tilde{F}_1(\mathbf{W}^{(i)}_{i},\mathbf{Z}^{(i)}_{k})-\frac{1}{2}\left\|\mathbf{W}_i\right\|^2_F-\frac{1}{2}\left\|\mathbf{Z}_k\right\|^2_F,\label{88}\\ 
	&\tilde{\mathbf{B}}^{(i)}_{k,l} \triangleq \tilde{F}_2(\mathbf{W}_k^{(i)},\mathbf{X}^{(i)}_{l})-\frac{1}{2}\left\|\mathbf{W}_k\right\|^2_F-\frac{1}{2}\left\|\mathbf{X}_l\right\|^2_F,\label{77}\\
	&\tilde{\mathbf{C}}^{(i)}_l \triangleq \tilde{F}_3(\mathbf{V}_{l}^{(i)},\mathbf{Y}^{(i)}_{l})-\frac{1}{2}\left\|\mathbf{V}_{l}\right\|^2_F-\frac{1}{2}\left\|\mathbf{Y}_l\right\|^2_F,\label{99}
	\end{align}
	respectively.~Then, we apply the following lemma to achieve a more efficient solution for the phase shifts, since the tightness of the SDR, i.e., $\text{Rank}(\mathbf{U}) = 1$ in (10g), cannot be ensured.
	\begin{lemma}
		The equivalent form of $\text{Rank}(\mathbf{U})= 1$, is given by \cite{Schoberr}
		\begin{equation}\label{rank}
		g(\mathbf{U})\triangleq\|\mathbf{U}\|_*-	\| \mathbf{U}\|_2 \leq 0.
		\end{equation}	
	\end{lemma}
	However,~(\ref{rank}) is still non-convex.~In order to tackle this obstacle,~we adopt a penalty approach to augment (15) into the objective function which penalizes the objective function when the matrix rank of $\mathbf{U}$ is greater than one.~Hence,~by using the first-order Taylor approximation of $\| \mathbf{U}\|_2$,~in each iteration of the MM
	algorithm, we obtain
	\vspace{-1.5mm}
	{\small\begin{align}\small \label{17}
		\tilde{g}^{(i)}(\mathbf{U})\triangleq&\|\mathbf{U}\|_*-\|\mathbf{U}^{(i)}\|_2-\text{Tr}\left[\mathbf{u}^{(i)}_{\max} (\mathbf{u}^{(i)}_{\max})^{\text{H}}(\mathbf{U}-\mathbf{U}^{(i)} )\right],
		\end{align}}\vspace{-1mm}where $\mathbf{u}^{(i)}_{\max}$ is the eigenvector corresponding to the maximum eigenvalue of matrix $\mathbf{U}^{(i)}$~in the $i$-th iteration.~Furthermore,~the non-convex constraint (\ref{p6-22c}) can be approximated as
	\vspace{-4mm}
	\begin{align}\label{18}
	\frac{\tilde{\mathbf{A}}^{(i)}_{k,k}}{\Gamma_{\text{req},k}}- \sum\limits_{ i \ne k}\hat{\mathbf{A}}^{(i)}_{k,i}-\sigma^2_{k}\geq 0,
	\end{align}
	where $\hat{\mathbf{A}}^{(i)}_{k,i}\triangleq\frac{1}{2}\left\|\mathbf{W}_i+\mathbf{Z}_k\right\|_F^2-\tilde{S}_1(\mathbf{W}^{(i)}_{i})-\tilde{S}_2(\mathbf{Z}^{(i)}_{k}),$ and
	\begin{align}\label{188}
	&S_1(\mathbf{W}_{i})\triangleq\frac{1}{2}\left\|\mathbf{W}_i\right\|^2_F\geq S_1(\mathbf{W}^{(i)}_{i})\nonumber\\&+\text{Tr}(\nabla_{\mathbf{W}_{i}}^HS_1(\mathbf{W}^{(i)}_{i})(\mathbf{W}_{i}-\mathbf{W}_{i}^{(i)}))\triangleq \tilde{S}_1(\mathbf{W}^{(i)}_{i}).
	\end{align}
	Similar to (\ref{188}), $\tilde{S}_2(\mathbf{Z}^{(i)}_{k})$ can be obtained by defining $S_2(\mathbf{Z}_{k})\triangleq\frac{1}{2}\left\|\mathbf{Z}_k\right\|^2_F$. As a result,~by augmenting (\ref{17}) to the objective function of (P5) with $\Phi\gg1$ as a penalty factor to penalize any non-rank-one matrix $\mathbf{U}$,~an upper bound of (P5) can be established via the following problem
	\begin{subequations}
		\begin{align}
		\text{P6:} \:\:
		&\underset{{\mathbf{V}},{\mathbf{W}}_k,{\mathbf{U}}} {\text{minimum}} \quad -\sum_{k\in\mathcal{K}}T_k +\Phi\big(\tilde{g}^{(i)}(\mathbf{U})\big) \\
		&\text{s.t.}\quad\eta_l\big( \sum_{l\in \mathcal{L}} \sum_{k\in \mathcal{K}} \tilde{\mathbf{B}}^{(i)}_{k,l}+ \sum_{l\in \mathcal{L}}\tilde{\mathbf{C}}^{(i)}_l\big) \geq\epsilon,\\
		&\quad\quad\eta_l \sum_{k\in \mathcal{K}} \tilde{\mathbf{B}}^{(i)}_{k,l}+\tilde{\mathbf{C}}^{(i)}_l  \ge    E_{\text{min},l},~ \forall l,\\
		&\quad\quad\text{(\ref{P6-0}),~(\ref{p6-5}),~(\ref{18})},~\text{Rank}(\mathbf{W}_k)= 1,
		\end{align}
	\end{subequations}
	where $T_k=	\log_2\big( \sum_{i \in \mathcal{K}} \tilde{\mathbf{A}}^{(i)}_{k,i}+ \sigma _k^2\big)-\tilde{R}(\mathbf{W}^{(i)}_{i},\mathbf{Z}^{(i)}_{k})$. Similar to (\ref{16}), $\tilde{R}(\mathbf{W}^{(i)}_{i},\mathbf{Z}^{(i)}_{k})$ denotes the lower bound of $	R(\mathbf{W}_{i},\mathbf{Z}_{k})\triangleq{\log_2\big( \sum_{i \in \mathcal{K},i \ne k} {{\mathbf{A}}_{k,i}}+ \sigma _k^2\big)}$.
	\begin{proposition}
		For arbitrary user channels,~the optimal solution to (P6) satisfies $\text{Rank}(\mathbf{W}^*_k)= 1$ and  $\mathbf{V}^*_l= \mathbf{0}$.   
	\end{proposition}
	\vspace{-2mm}
	\textit{Proof.}~It can be proved by following a similar approach as in \cite{Wu5}, which is omitted here due to page limitation.
	
	
	It can be observed that (P6) is a convex optimization problem and optimization tools such as CVX can be utilized to solve it efficiently\cite{Grant}.~By iteratively solving (P6) optimally, we can monotonically tighten this upper bound. Besides, the objective function in (P6) is monotonically non-increasing, which guarantees the converge to a stationary point.~The maximum value of $\epsilon$ is obtained with ${E}_\text{max}$ such that (P5) remains feasible\cite{Khalili},~where ${E}_\text{max}$ is the maximum EH amount.~The value of ${E}_\text{max}$ can be found by solving the following optimization problem:
	\begin{subequations}
		\begin{align}
		\text{P7:} \:\: &\underset{{\mathbf{V}},{\mathbf{w}}_k,{\mathbf{\Theta}}} {\text{maximum}} \quad  \sum_{l\in \mathcal{L}} P_l({\mathbf{w}}_k,{\mathbf{V}},{\mathbf{\Theta}})-\Phi\big(\tilde{g}^{(i)}(\mathbf{U})\big)\\
		&\text{s.t.}\quad \eta_l \sum_{k\in \mathcal{K}} \tilde{\mathbf{B}}^{(i)}_{k,l}+\tilde{\mathbf{C}}^{(i)}_l  \ge    E_{\text{min},l},~ \forall l,\\
		&\quad\quad\text{(\ref{P6-0}),~(\ref{p6-5}),~(\ref{18})},~\text{Rank}(\mathbf{W}_k)= 1.
		\end{align}
	\end{subequations}
	The different values of $\epsilon$ lead to different trade-offs between total EH amount and data sum-rate.~To obtain a specific value of $\epsilon$,~we let $\epsilon= \delta {E}_\text{max},$ where $\delta$ is a positive value in the range of $(0,~1]$.
	\vspace{-5mm}
	\subsection{Computational Complexity Analysis}
	In this subsection, we present a computational complexity for our proposed solution.~Specifically,~(P6) includes $M^{2}$ variables and $3K$ affine constraints.~Consequently, the complexity order for designing joint beamforming and phase shift optimization in each iteration is given by $\mathcal{O}\big(\log\frac{1}{\epsilon}(3K)M^{2}+N^{2}\big)^{3.5}$\cite{Complexity}.~Moreover,~the computational complexity order for solving (P6) via adopting AO method based on the SDP for finding beamforming is $\mathcal{O}\big(M^{2}+3K\big)^{3.5}$ while for the reflecting elements the complexity order is $\mathcal{O}(3K+N^{2})^{3.5}$\cite{Complexity}.
	It is worth mentioning that while the KKT solution obtained by the IA algorithm is better in quality than the stationary point obtained by the AO algorithm, there is a trade-off between the algorithm complexity and the system performance.
	\vspace{-4mm}
	\section{Simulation Results}
\begin{figure}[t]
		\centering
		\includegraphics[width=2.50in] {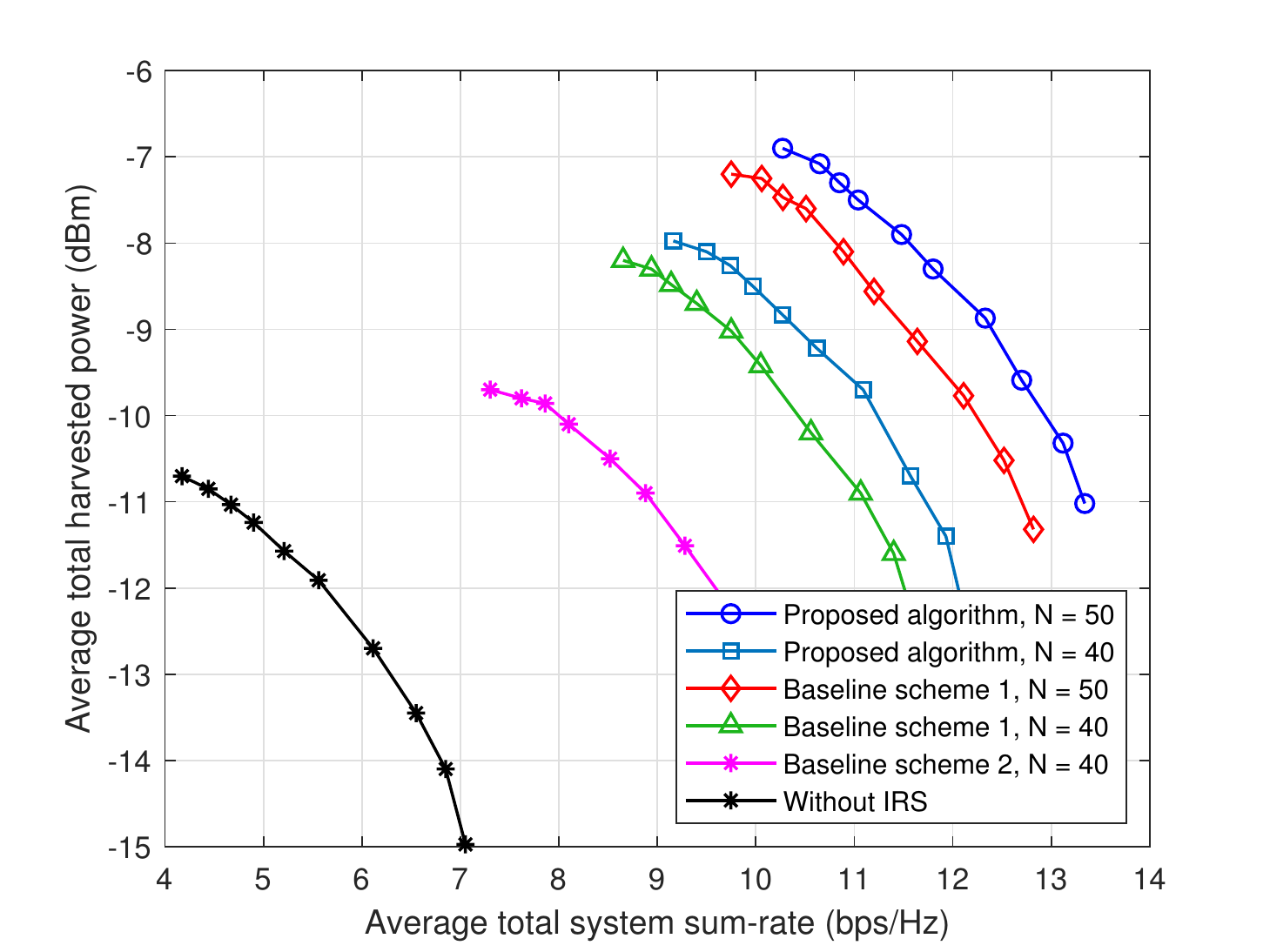}
		\caption{\small Rate-energy trade-off region.}\label{fig1}
		\vspace{-7mm}
	\end{figure}

	We evaluate the performance of the proposed algorithm by simulation.~The simulation parameters are summarized in Table \ref{table-notations} unless otherwise is given.~We consider the location of the BS and the IRS as $(3,0)$ m and $(0,4)$ m,~respectively.~Also,~it is assumed that $K=2$ IDRs and $L=2$ EHRs are randomly distributed with a distance of $d=50$ m and $d=4$ m from the BS,~respectively.~The small-scale fading channels are modeled as Rayleigh fading.
	\begin{table}[t]
		\renewcommand{\arraystretch}{1.05}
		\centering
		\caption{Simulation Parameters}
		\label{table-notations}
		\begin{tabular}{| c| c| }    
			\hline
			\textbf{Parameters}& \textbf{Values}\\\hline        
			Path-loss model and exponent& \cite{Wu5}\\ \hline
			Number of antennas at the BS,~$M$ & $4$\\ \hline
			Maximum transmit power,~$P_{\text{max}}$ & $40$ dBm\\ \hline
			Minimum required SINR,~$\Gamma_{\text{req},k}$ & $5$ dB\\ \hline
			Minimum required EH,~$E_{\text{min},l}$ &  $-20$ dBm\\ \hline
			Carrier frequency& $750$ MHz\\ \hline
			Noise power,~$\sigma _k^2$ & $\sigma_k^2=\sigma^2=-90~\text{dBm}$  \\ \hline
		\end{tabular}
		\vspace{-5mm}
	\end{table}
	Fig.~\ref{fig1} investigates the trade-off region between the system data sum-rate and average total harvested energy for different values of $N$,~which is obtained via solving (P6) by varying the values of $\delta$ with a step size of $0.1$.~As can be observed the average harvested energy decreases with the increasing data sum-rate.~This result confirms that the objective of maximizing average harvested energy generally conflicts with that of maximizing the data sum-rate.~For comparison,~we also consider two baseline schemes.~For baseline scheme 1,~we consider the proposed scheme in \cite{Wu5},~where SDP and AO approaches are adopted and apply them with the $\epsilon$-constraint method for the considered MOOP framework.~For baseline scheme 2,~we consider random passive beamforming at the IRS \cite{Wu2}.~It can be observed that our proposed scheme can establish a better performance compared with baseline scheme 1.~Note that in our proposed solution, we optimize the active beamforming at the AP and passive beamforming at the IRS simultaneously in every iteration, while the proposed algorithm in the literature adopted an AO method only optimizes parts of the total variables each time. In particular, the objective function of (P6) is non-decreasing in each iteration and the proposed IA algorithm is guaranteed to converge to a KKT solution \cite{Schoberr}. Hence, the superior performance brought by the proposed scheme is due to the fact that the AO algorithm can be easily trapped in some inefficient solution, while the IA algorithm can ensure the convergence to a KKT solution of the design problem, which is better in quality than the point obtained by the AO algorithm in general.~In addition,~our proposed scheme also outperforms baseline scheme 2 without phase-shift optimization.~On the other hand, one can observe that by increasing the number of IRS reflecting elements, $N$, a larger achievable trade-off region can be obtained, which demonstrates the superiority of deploying IRS with a large number of low-cost reflecting elements.~We can also observe that the achieved average trade-off region is significantly enlarged by deploying an IRS in the considered system. This confirms the capability of the IRS as a promising approach for providing favorable channel conditions, which is beneficial to the data sum-rate as well as harvested energy.
	\begin{figure}[t]
		\centering
		\includegraphics[width=2.50in] {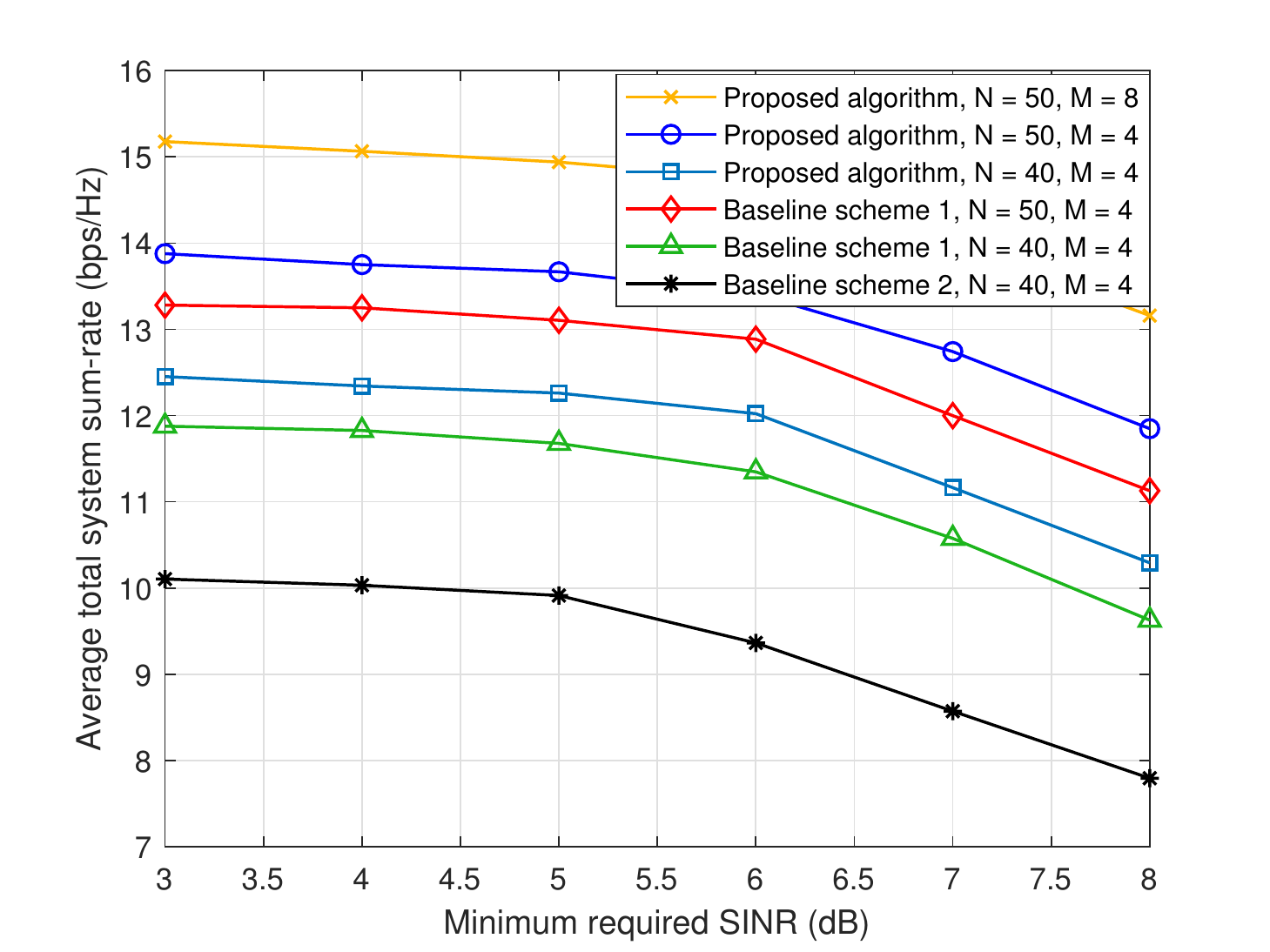}
		\caption{\small Average sum-rate versus minimum SINR.}\label{fig2}
		\vspace{-7.5mm}
	\end{figure}Fig.~\ref{fig2} plots the sum-rate versus the minimum target SINR ($\Gamma_{\text{req}}$) for different values of $M$ and $N$,~which is obtained by solving (P6) with $\epsilon=0$.~It can be perceived that the data sum-rate remains nearly constant for a small value of minimum required SINR,~$\Gamma_{\text{req}}$,~but starts to decline as $\Gamma_{\text{req}}$ increases.~This is because when $\Gamma_{\text{req}}$ is low,~the transmitted power budget is low.~Therefore,~the proposed design can easily satisfy the constraints.~However,~for a higher value of $\Gamma_{\text{req}}$,~more transmit power is required to meet the target SINR.~It can be seen that our proposed scheme outperforms the two baseline schemes, which shows the effectiveness of the proposed design based on the IA-method for jointly optimizing the reflecting elements and active beamformers at the IRS and the BS, respectively. We also observe the effect of increasing transmit antennas at the BS, $M$, as well as reflecting elements at the IRS, $N$, on the performance gain in terms of average sum-rate. In particular, increasing $M$ and $N$ allows a further improvement of the average total system sum-rate achieved by the proposed design. More specifically, additional reflecting elements at the IRS offer more degrees of freedom for resource allocation by establishing a more favorable propagation environment.~Furthermore, increasing the number of transmitting antennas at the BS provides a higher spatial multiplexing gain, which results in an improvement of average sum-rate.
	\vspace{-3.5mm}
	\section{Conclusion}
	\vspace{-1.5mm}
	In this letter,~an MOOP was formulated for the joint passive and active beamforming design in an IRS-aided SWIPT system to study the trade-off between the data sum-rate maximization and the total harvested energy maximization.~We first applied the $\epsilon$-constraint method to convert the MOOP into an SOOP and then proposed an IA-based algorithm to obtain an efficient solution.~Simulation results unveiled the advantages of the IRS deployment and demonstrated the superior performance of the proposed scheme as compared with baseline schemes.

\end{document}